\documentclass[reprint,amsmath,amssymb,pra]{revtex4-1}
\usepackage{amsmath,epsfig,wrapfig,graphicx,color,bm,epstopdf}

\begin{document}
\title{Chemical composition induced quantum phase transition in 
 Cs$_{1-x}$Rb$_{x}$FeCl$_{3}$}

\author{S. Hayashida,$^{1}$ L. Stoppel,$^{1}$ Z. Yan,$^{1}$ S. Gvasaliya,$^{1}$ A. Podlesnyak,$^{2}$ and A. Zheludev$^{1}$}
\affiliation{$^{1}$Laboratory for Solid State Physics, ETH Z{\"u}rich, 8093 Z{\"u}rich, Switzerland \\
$^{2}$Neutron Scattering Division, Oak Ridge National Laboratory, Oak Ridge, Tennessee 37831, USA}

\date{\today}

\begin{abstract}
The isostructural series of $S=1$ quantum magnets Cs$_{1-x}$Rb$_{x}$FeCl$_{3}$ is investigated, using both thermodynamic measurements and inelastic neutron scattering experiments.
It is found that increasing with Rb content the system evolves from the gapped state at $x=0$, through a quantum phase transition at $x\sim 0.35$, and to the magnetically ordered state at larger $x$. Inelastic neutron experiments for $x=0$, $x=0.3$, and $x=1$ demonstrate that the magnetic anisotropy and spin interactions are continuously tuned by the chemical composition. For the intermediate concentration all magnetic excitations are substantially broadened suggesting that disorder plays a key role in this species. For the two end compounds, excitations remain sharp. 
\end{abstract}

\maketitle

\section{Introduction}\label{sec1}
Gapped quantum paramagnets are an excellent platform for studying quantum phase transitions (QPTs).
In these systems QPTs leading to a magnetically ordered state can be induced by magnetic field, pressure, or chemical modification~\cite{Sachdev2000,sachdev2007quantum,Sachdev2008}.
Field-induced QPTs of this type have been extensively studied in numerous compounds, a review of 
which can be found in Ref.~\cite{Zapf2014}.
Pressure-induced QPTs have been also reported in several systems including TlCuCl$_{3}$~\cite{Tanaka2003,Ruegg2004,Ruegg2008}, 
piperazinium-Cu$_{2}$Cl$_{6}$~\cite{Thede2014,Perren2015,Mannig2016}, 
and CsFeCl$_{3}$~\cite{Kurita2016,Hayashida2018}.
Recent attention has been drawn to the less common chemical composition induced QPTs in quantum magnets~\cite{Vojta2013,Zheludev2013}.
Unlike the field- and pressure-induced QPTs, in composition-driven QPTs disorder effects are endemic because of a local structural-distortion by the chemical modification. 
The presence of such disorder can change the nature of the quantum critical point (QCP)~\cite{Zheludev2013}, or even destroy it entirely~\cite{Vojta2013}.
QPTs in the presence of disorder are complex phenomena, understanding which can benefit from further experimental input. 

The effect of the chemical substitution has actually already been studied in most known quantum magnets, including 
Tl$_{1-x}$K$_{x}$CuCl$_{3}$~\cite{Oosawa2002,Shindo2004}, 
(CH$_{3}$)$_{2}$CHNH$_{3}$Cu(Cl$_{1-x}$Br$_{x}$)$_{3}$~\cite{Nafradi2013,Perren2018}, 
piperazinium-Cu$_{2}$(Cl$_{1-x}$Br$_{x}$)$_{6}$~\cite{Huvonen2012_1,Huvonen2012_2,Huvonen2013,Glazkov2014}, 
H$_{8}$C$_{4}$SO$_{2}\cdot$Cu$_{2}$(Cl$_{1-x}$Br$_{x}$)$_{4}$~\cite{Wulf2011}, 
and Cu(quinoxaline)(Cl$_{1-x}$Br$_{x}$)$_{2}$~\cite{Keith2011,Povarov2014}. 
Unfortunately, all these systems are pushed {\em away} from criticality by such modification. 
To date, the only known composition-induced QPT is in a gapped quantum magnet Ni(Cl$_{1-x}$Br$_{x}$)$_{2}\cdot$4SC(NH$_{2}$)$_{2}$ (DTNX),
where disorder effects do not seem to play a huge role~\cite{Yu2012,Wulf2013,Povarov2015,Povarov2017,Mannig2018}. In the present  work we present another example, where disorder effects are much more prominent.

\begin{figure}[tbp]
	\includegraphics[scale=1]{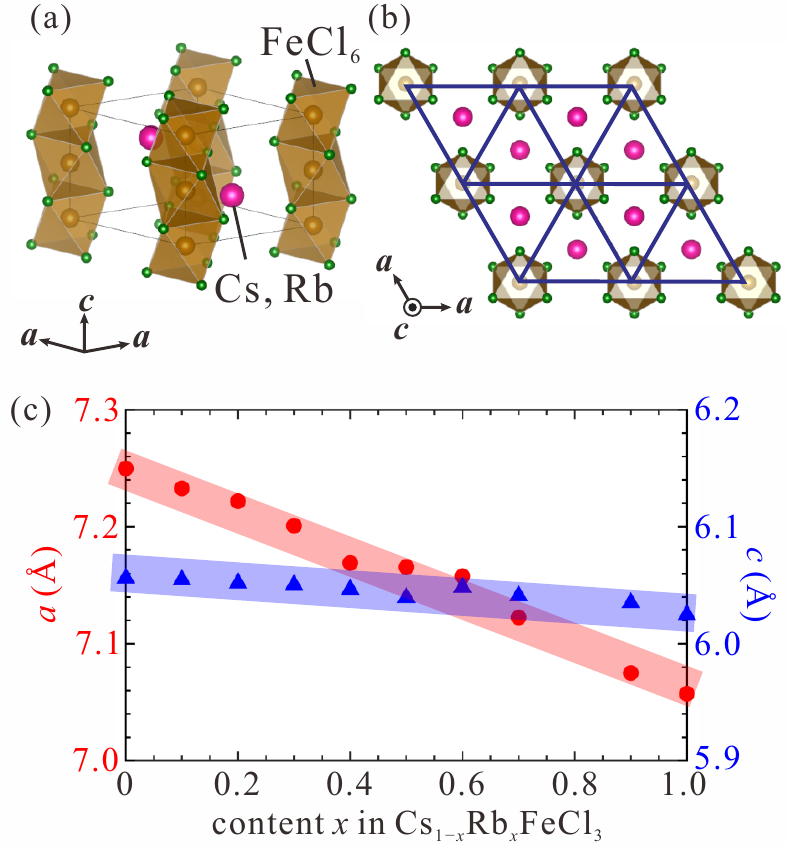}
	\caption{(a) Crystal structure of Cs$_{1-x}$Rb$_{x}$FeCl$_{3}$ with space group $P6_{3}/mmc$.
(b) Triangular lattice in the $ab$ plane. 
(c) Measured lattice constants for all compounds studies in this work. Red circles and blue triangles indicate lattice constants $a$ and $c$, respectively. Shaded areas are guides for eye.}
	\label{fig1}
\end{figure}

Our target compound is an easy-plane type antiferromagnet Cs$_{1-x}$Rb$_{x}$FeCl$_{3}$. 
Species of this isostructural series crystallize in a hexagonal structure with space group $P6_{3}/mmc$ as displayed in Figs.~\ref{fig1}(a) and \ref{fig1}(b).
Straightforward powder x-ray diffraction measurements yield lattice constants $a=7.2499(1)$~{\AA} and $c=6.0561(1)$~{\AA} for CsFeCl$_{3}$ and
$a=7.0572(1)$~{\AA} and $c=6.0247(1)$~{\AA} for RbFeCl$_{3}$, in agreement with previous studies~\cite{Seifert1966}.
For intermediate Rb content the lattice parameters change continuously as shown in Fig.~\ref{fig1}(c).

The magnetic properties of Cs$_{1-x}$Rb$_{x}$FeCl$_{3}$ are due to Fe$^{2+}$ ion ($3d^{6}$, $S=2$, $L=2$).
The FeCl$_{6}$ octahedra form one-dimensional chains along the crystallographic $c$ axis [Fig.~\ref{fig1}(a)], 
and a triangular structure in the $ab$ plane [Fig.~\ref{fig1}(b)].
The low-energy excitation of the Fe$^{2+}$ ion is described by a pseudo-spin $S=1$ due to the cubic
crystal field and spin-orbit coupling~\cite{Inomata1967,Eibschutz1975}.
The spin system has been identified as ferromagnetic chains.
These chains are weakly coupled by antiferromagnetic interaction in the triangular plane~\cite{Montano1973,Yoshizawa1980,Steiner1981}.
Since they have strong easy-plane type anisotropy, the system is regarded as $S=1$ easy-plane type
antiferromagnet~\cite{Matsumoto2007}.
The easy-plane anisotropy splits the triplet spin $S=1$ into the singlet $S^{z}=0$ and the doublet
$S^{z}=\pm 1$, favoring a gapped non-magnetic ground state which is a quantum disordered (QD) phase.
In contrast, the spin interaction favors a magnetic long-range ordered (LRO) state.
Controlling the competition between the anisotropy and the spin interaction by an external parameter 
brings about the QPT.

CsFeCl$_{3}$ exhibits a gapped non-magnetic ground state with $S^{z}=0$ on each site~\cite{Yoshizawa1980}, and a pressure-induced QPT~\cite{Kurita2016,Hayashida2018}.
 In contrast, RbFeCl$_{3}$ undergoes a three magnetic transitions at $T_{\rm N1}=2.5$~K, $T_{\rm N2}=2.35$~K, and $T_{\rm N3}=1.95$~K~\cite{Haseda1981,Wada1983}. 
The ground state below $T_{\rm N3}=1.95$~K is a 120$^{\circ}$ structure having the propagation vector $(1/3,1/3,0)$~\cite{Wada1983}.
In this paper we show that, similarly to what is seen in  DTNX~\cite{Povarov2015,Povarov2017,Mannig2018}, the spin gap in Cs$_{1-x}$Rb$_{x}$FeCl$_{3}$  closes at some critically value $x=x_{c}$, driving a QPT and eventually restoring magnetic long-range order.

\section{Experimental details}\label{sec2}
Single-crystal samples were grown by the vertical Bridgman method~\cite{Kurita2016}. 
The crystals were aligned using the Bruker APEX-II single crystal x-ray diffractometer.
Bulk measurements were carried out using the Quantum Design Physical Property Measurement System (PPMS).
Heat capacity was measured on a standard Quantum Design relaxation calorimetry 
option and the $^{3}$He-$^{4}$He dilution refrigerator insert for PPMS. 
We measured temperature scans at zero magnetic field and field scans at 0.25 K for each sample.
Field dependence of alternating current (ac) magnetic susceptibility was measured at 2 K.
In both measurements of the heat capacity and ac magnetic susceptibility, magnetic field was applied along the $c$ axis.

Inelastic neutron scattering (INS) experiments were performed at the Cold Neutron Chopper Spectrometer (CNCS)~\cite{cncs2011,cncs2016} at the Spallation Neutron Source (SNS) of the Oak Ridge National Laboratory in USA.
Three samples were investigated: the parent compounds CsFeCl$_{3}$ and RbFeCl$_{3}$, as well as the $x=0.3$ material Cs$_{0.7}$Rb$_{0.3}$FeCl$_{3}$.
In each case the samples were aligned such that corresponding $(a^{*}+b^{*})$-$c^{*}$ planes were horizontal.
All measurements were performed at 0.1~K maintained by a $^{3}$He-$^{4}$He dilution cryostat. 
Data were taken with incident neutron beam energies $E_{\rm i}=5.93$ or 2.99 meV. The energy resolution at elastic position was $\Delta E=0.23$ and 0.09 meV, respectively.
For each incident energy, the samples were rotated stepwise to fully cover the spectra in the scattering plane.
In the following, all spectra shown represent data integrated in momentum transfer perpendicular to the horizontal plane in the range $|q_{\perp}| \leq 0.05$~{\AA}$^{-1}$.

\section{Results and discussion}\label{sec3}
\subsection{Thermodynamics}

\begin{figure}[bp]
	\includegraphics[scale=1]{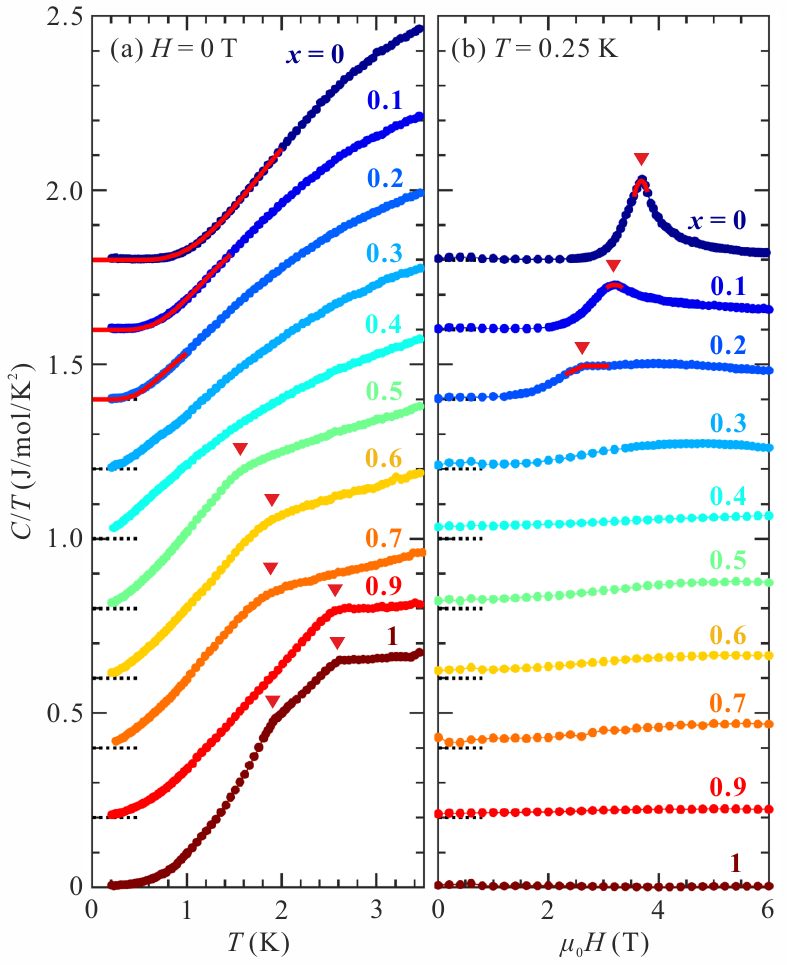}
	\caption{Heat capacity measurements for all compounds. (a) Temperature scans at zero magnetic field and (b) field scans in an applied magnetic field along the $c$ axis at $T=0.25$~K. 
	$x$ is the content of the Rb. 
The solid curves in (a) are the fits proportional to $\exp(-\Delta/k_{\rm B}T)$.
The solid curves and lines in (b) are empirical fits to estimate the positions of peaks and kinks. 
These are indicate by red triangles.
The data are shifted by vertical offset of 0.2.
}
	\label{fig2}
\end{figure}

\begin{figure}[tbp]
	\includegraphics[scale=1]{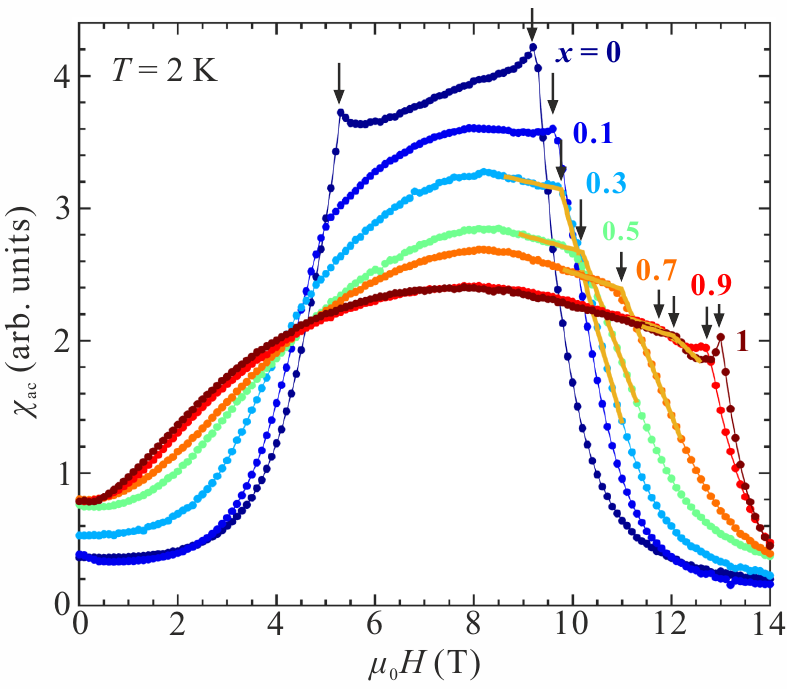}
	\caption{Magnetic susceptibility at $T=2$ K in an applied magnetic field for selected compounds.  The magnetic field is applied along the $c$ axis.
The solid lines are empirical fits to estimate positions of the kinks. 
The red triangles indicate positions of the peaks and kinks.}
	\label{fig3}
\end{figure}

The measured temperature dependencies of  heat capacity in zero magnetic field are shown in Fig.~\ref{fig2}(a).
For the parent compound CsFeCl$_{3}$, the specific heat below 2~K has a pronounced activated form 
$\exp(-\Delta/k_{\rm B}T)$, as indicated by the red solid curve.
The activation energy $\Delta$ is determined to be 0.529(1) meV.
The low temperature heat capacities for $x=0.1$ and $0.2$ are also reasonably well described by activation laws with activation energies are 0.346(2) and 0.225(3) meV, respectively.
We conclude that the spin gap decreases with increasing $x$ for small $x$.

On the Rb side, two kinks are observed at $T=2.6$, and 1.9 K for the parent compound RbFeCl$_{3}$ as indicated by the triangles in Fig.~\ref{fig2}(a).
These correspond to the transitions at $T_{\rm N1}$ and $T_{\rm N3}$, as reported in Refs.~\cite{Haseda1981,Wada1983}.
At around $T=1$ K there is a weak indication of an additional feature in the temperature dependence for the $x=1$ material. 
However, repeated measurements on several samples were unable to unambiguously clarify its significance.
For $x=0.9$, only the second kink at higher temperature is visible.  
Single kinks indicative of long-range ordering are also observed at $x=0.5, 0.6$, and $0.7$, but not at $x=0.4$.
The position of the kink decreases with decreasing the Rb content, indicating that the transition 
temperature reduces upon approaching the critical point.
We conclude that  $x=0.4$ is close to being at the critical concentration.

The field scans of the heat capacity at 0.25 K are shown in Fig.~\ref{fig2}(b).
A sharp peak is observed at $H_{c1}=$ 3.7 T for $x=0$, which corresponds to a phase transition from the gapped state to the antiferromagnetically ordered one.
The transition field is consistent with the previous reports~\cite{Kurita2016}.
A peak and kink are also observed for $x=0.1$ and 0.2.
Their positions are shown in Fig.~\ref{fig2}(b) by red triangles and visibly shift towards lower field with increasing $x$.
This is also consistent with a decrease of the spin gap. The peak in specific heat becomes progressively broader and disappears at $x=0.3$.
Overall, the observed broadening of the heat capacity peak is much more pronounced than in the similar system DTNX~\cite{Povarov2017}.
This suggests that in Cs$_{1-x}$Rb$_{x}$FeCl$_{3}$ the magnetic state is strongly affected by disorder caused by the chemical substitution.


As shown in Fig.~\ref{fig3}, the magnetic susceptibility of Cs$_{1-x}$Rb$_{x}$FeCl$_{3}$ systematically evolves as a function of Rb concentration $x$.
For $x=0$, sharp anomalies are observed at the field of gap closure $H_{c1}=5.2$ T and at saturation $H_{c2}=9.2$~T. 
For $x=1$ the saturation field is $H_{c2}=13$ T.
The anomaly at lower fields is rapidly broadened with increasing Rb content, but pronounced kinks near saturation persist at all concentrations.
This implies that disorder affects the low-energy gap excitations more than the high-energy zone-boundary states.
For $x=1$ and 0.9, there are two separate features near $H_{c2}$.
The kinks at about 12 T continue the trend found in lower Rb content materials, and another sharp feature appears at about 13 T.

\begin{figure}[tbp]
		\includegraphics[scale=1]{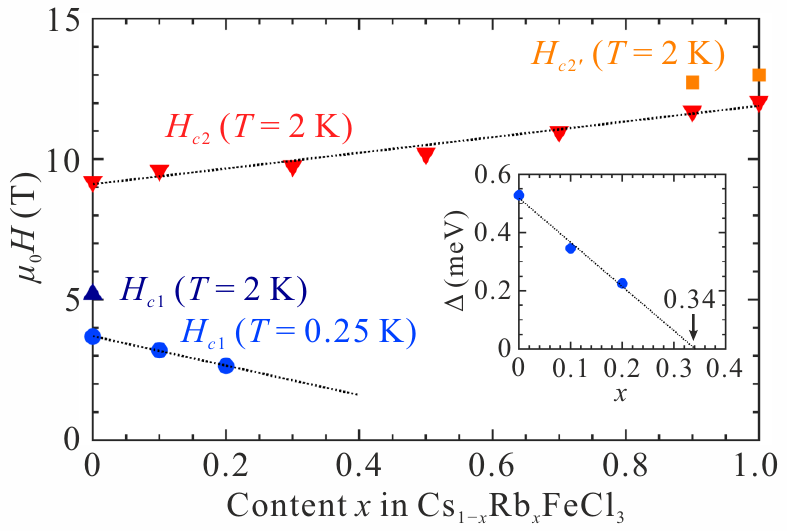}
	\caption{Transition fields $H_{c1}$, $H_{c2}$ plotted as a function of the Rb content in Cs$_{1-x}$Rb$_{x}$FeCl$_{3}$, as determined from heat capacity and ac magnetic susceptibility measurements. The inset shows the behavior of the activation
energy in the zero-field heat capacity experiments. The dashed solid lines are linear fits.
The estimated error bars are inside the data symbols.
}
	\label{fig4}
\end{figure}

The measured critical fields $H_{\rm c1}$ and  $H_{\rm c2}$ are plotted as a function of the Rb content in
Fig.~\ref{fig4}.
The resulting phase diagram indicates that the concentration $x$ systematically tunes the spin Hamiltonian.
The concentration evolution of the gap energies estimated from the heat capacity
is shown in the inset of Fig.~\ref{fig4}.
From a linear extrapolation, the critical concentration can be estimated as $x_{c}\sim 0.35$.

\subsection{Inelastic neutron scattering}
\begin{figure}[tbp]
		\includegraphics[scale=1]{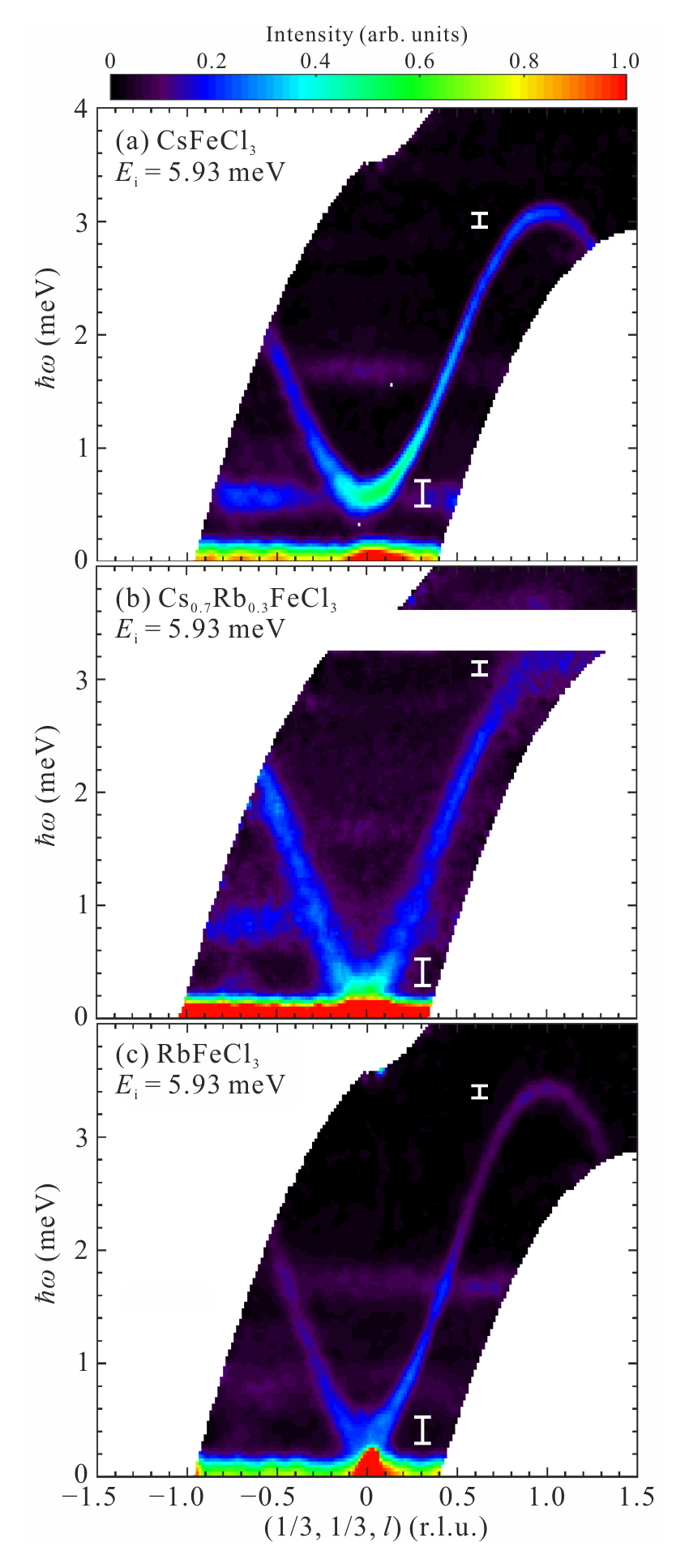}
	\caption{False color plots of the measured inelastic neutron intensity projected onto the $\hbar\omega$-$(1/3,1/3,l)$ plane for 
(a) CsFeCl$_{3}$, (b) Cs$_{0.7}$Rb$_{0.3}$FeCl$_{3}$, and (c) RbFeCl$_{3}$ at $T\simeq 0.1$ K.
The incident neutron energy is $E_{\rm i}=5.93$ meV.
The spectra are integrated in the range of $0.30(a^{*}+b^{*})\leq q \leq 0.36(a^{*}+b^{*})$.
The white bars indicate the calculated instrument energy resolution.
The white area at $\hbar\omega = 3.5$~meV in (b) masks a strong spurious scattering that does not originate from the sample.
}
	\label{fig5}
\end{figure}

\begin{figure}[tbp]
	\includegraphics[scale=1]{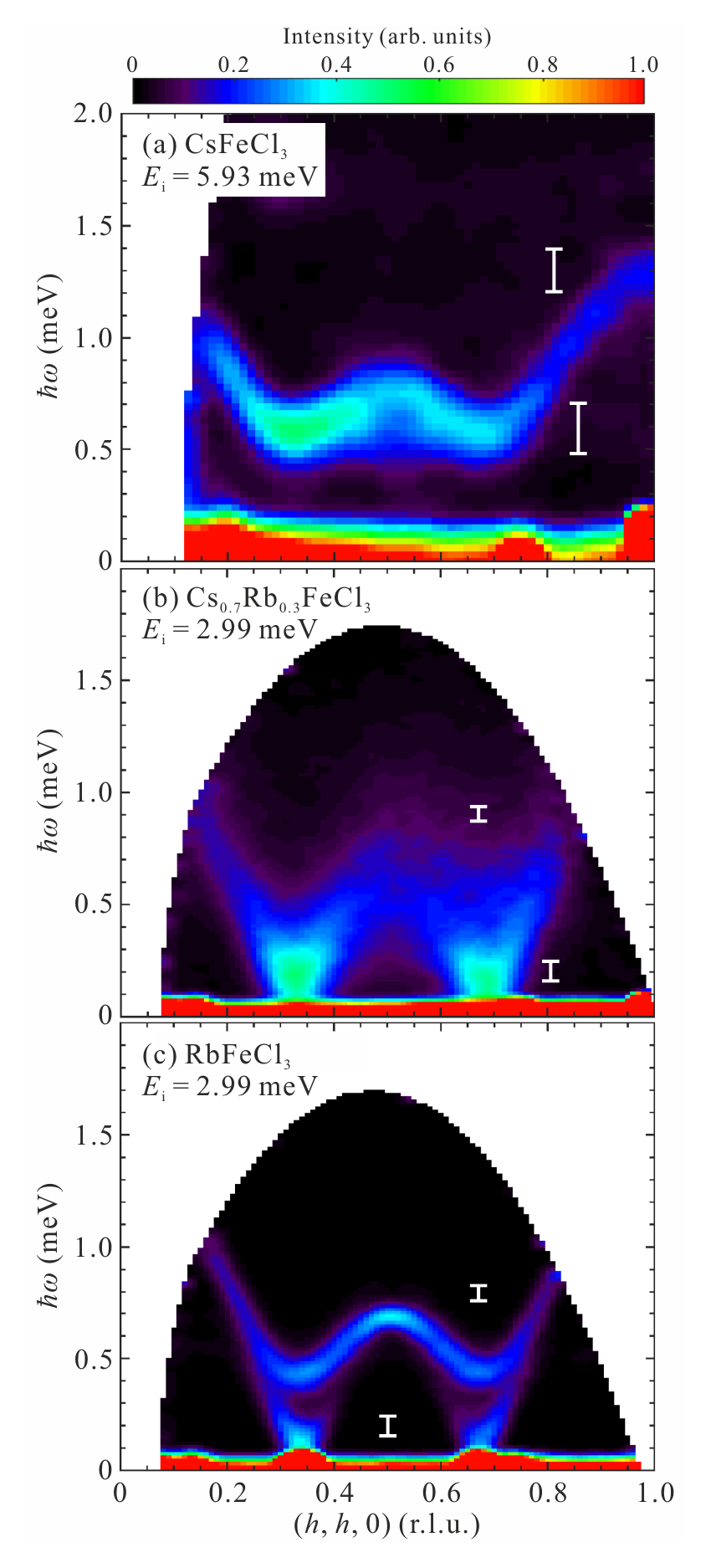}
	\caption{False color plots of the measured inelastic neutron intensity projected onto the $\hbar\omega$-$(h,h,0)$ plane for (a) CsFeCl$_{3}$,
(b) Cs$_{0.7}$Rb$_{0.3}$FeCl$_{3}$, and (c) RbFeCl$_{3}$ at $T\simeq 0.1$ K.
The incident neutron energies are $E_{\rm i}=5.93$ meV for CsFeCl$_{3}$, and $E_{\rm i}=2.99$ meV for Cs$_{0.7}$Rb$_{0.3}$FeCl$_{3}$ and RbFeCl$_{3}$.
The spectra are integrated in the range of $-0.05c^{*}\leq q \leq 0.05c^{*}$. The white bars indicate the instrumental energy resolutions.}
	\label{fig6}
\end{figure}

The measured INS spectra for CsFeCl$_{3}$, Cs$_{0.7}$Rb$_{0.3}$FeCl$_{3}$ and RbFeCl$_{3}$  are visualized in Figs.~\ref{fig5}
 and \ref{fig6} as cuts along the $c^{*}$ 
 and $a^{*}+b^{*}$ directions, respectively.
They are presented as false color plots of the neutron intensity, and are plotted without any background
subtraction.
The flat intensity bands at $\hbar\omega=0.8$ and 1.8 meV in Figs.~\ref{fig5}(a)-\ref{fig5}(c) are spurious
and have been observed to shift their position when the incident energy is changed.
They are likely due to multiple scattering in the sample environment.
Note that different panels show measurements with two different incident energies and thus different energy resolution, as indicated by the white vertical bars in the figures.

\begin{figure}[tbp]
		\includegraphics[scale=1]{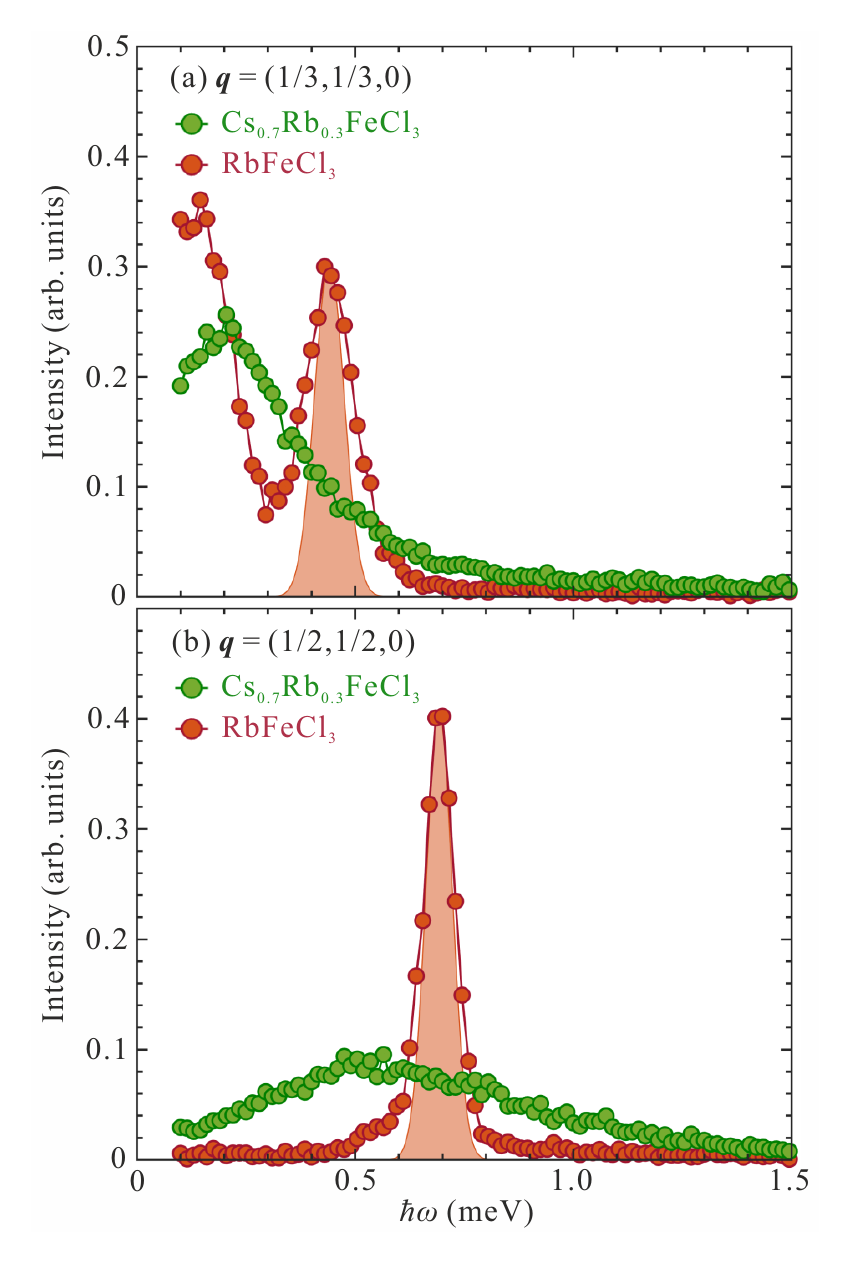}
	\caption{Constant-${\bm q}$ cuts at (a) $(1/3,1/3,0)$ and (b) $(1/2,1/2,0)$ 
for Cs$_{0.7}$Rb$_{0.3}$FeCl$_{3}$ and RbFeCl$_{3}$. 
The spectra were integrated  in the range of $0.30(a^{*}+b^{*})\leq q \leq 0.36(a^{*}+b^{*})$ and $-0.05c^{*}\leq q \leq 0.05c^{*}$ for (a), and $0.47(a^{*}+b^{*})\leq q \leq 0.53(a^{*}+b^{*})$ and $-0.05c^{*}\leq q \leq 0.05c^{*}$ for (b).
The shaded areas represent the calculated instrumental energy resolution.}
	\label{fig7}
\end{figure}

\begin{figure}[tbp]
	\includegraphics[scale=1]{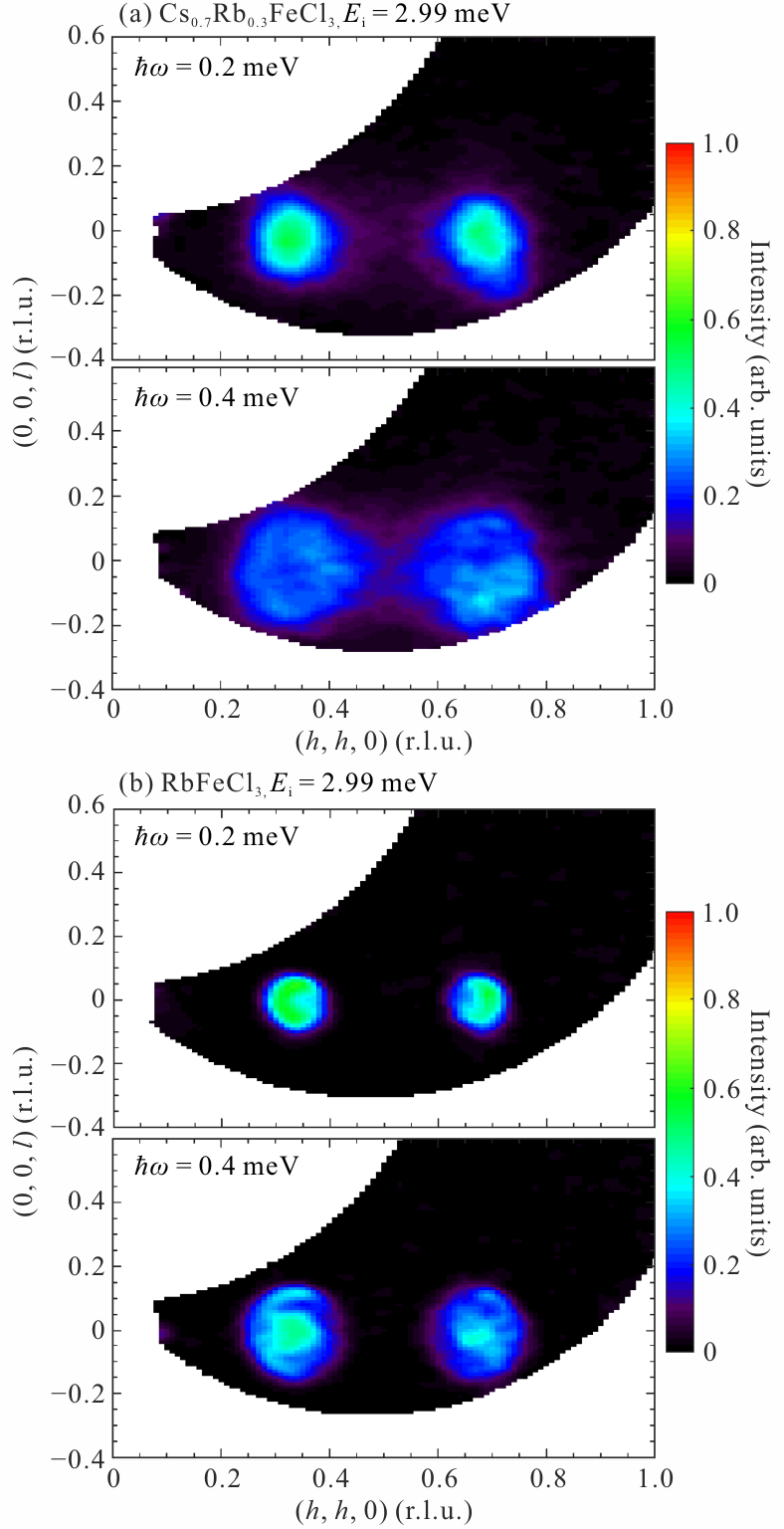}
	\caption{False color plots of inelastic neutron intensity measured in (a) Cs$_{0.7}$Rb$_{0.3}$FeCl$_{3}$and (b) RbFeCl$_{3}$ at constant energy transfers of 0.2 meV and 0.4 meV.
		The data were taken with $E_{\rm i}=$ 2.99~meV. 
The spectrum was integrated in the range of 0.175 meV $\leq \hbar\omega \leq$ 0.225 meV and 
0.375 meV $\leq \hbar\omega \leq$ 0.425 meV. The aspect ratio
of the figure is scaled to units of {\AA}$^{-1}$.}
	\label{fig8}
\end{figure}

Highly dispersive magnetic excitations are clearly observed below $\hbar\omega=4$ meV along the $c^{*}$ direction and below 1.5 meV along the $a^{*}+b^{*}$ direction in all samples. 
As borne out in  Figs.~\ref{fig5}(a) and \ref{fig6}(a), in CsFeCl$_{3}$ there is a single excitation branch with a gap of 0.59 meV in full agreement with previous neutron studies ~\cite{Yoshizawa1980} and with the activation energy of the heat capacity. 
In RbFeCl$_{3}$ our experiments show $(1/3,1/3,0)$ magnetic Bragg peaks appearing at low temperatures. 
The magnetic propagation vector is consistent with  previous  neutron diffraction study~\cite{Wada1983}. 
However, unlike the previously reported lower-resolution inelastic experiments~\cite{Yoshizawa1980}, {\em two} distinct  spin-wave branches are observed in the inelastic channel. 
One of these has a gap of 0.5 meV at $(1/3,1/3,0)$, while the other is gapless (see Fig.~\ref{fig6}(c)).

The most interesting results pertain to Cs$_{0.7}$Rb$_{0.3}$FeCl$_{3}$. 
Only a single mode is visible. 
However, compared to the $x=0$ parent compound, the bandwidth is increased and the spin gap is almost closed, as shown in Fig.~\ref{fig5}(b).
No elastic (Bragg) scattering was observed at the wave vector of the dispersion minimum, suggesting that the system is still in the quantum paramagnetic phase.
Figures~\ref{fig7} show constant-${\bm q}$ cuts at ${\bm q}=(1/3,1/3,0)$ and $(1/2,1/2,0)$.
Whether or not there is a true gap at the former reciprocal space position is not clear cut, although the observed  broad peak is centered at a finite energy of 0.2 meV.
We conclude that the concentration of $x=0.3$ is close to critical. 
Constant energy slices below 0.4 meV are shown in Fig.~\ref{fig8}(a).
These are contained in a well-defined and almost isotropic ``relativistic'' cone, suggesting that the proximate QCP is a three-dimensional one with a dynamical exponent $z=1$, similarly to DTNX~\cite{Povarov2015,Povarov2017,Mannig2018}.
Compared to the $x = 1$ parent compound in Fig.~\ref{fig8}(b), the spectrum in the $x = 0.3$ compound broadens in the ${\bm q}$ space and loses its distinct concentric structure.

A key finding of this work is that  the magnetic excitations in Cs$_{0.7}$Rb$_{0.3}$FeCl$_{3}$ are obviously broadened compared to those in the parent compounds. 
Note that the data shown in Figs.~\ref{fig7} were taken with the {\em high} resolution setup $E_{\rm i}=2.99$~meV. 
The intrinsic width of the measured inelastic features is also apparent in the scans plotted in Figs.~\ref{fig7}. 
We attribute the broadening effect to disorder, as in  DTNX~\cite{Povarov2017,Mannig2018}. 
Note, however, that the relative broadening appears to be considerably more pronounced in  Cs$_{0.7}$Rb$_{0.3}$FeCl$_{3}$. 
Unfortunately, this extreme broadening in  Cs$_{0.7}$Rb$_{0.3}$FeCl$_{3}$, as well as the presence of multiple and often poorly resolved branches in RbFeCl$_{3}$, prevents us from carrying out a consistent quantitatively spin-wave theoretical analysis of the excitation spectra, as it was done for DTNX~\cite{Mannig2018}.

\section{Conclusion}
The qualitative conclusions of this study are rather unambiguous.  For $x\sim 0.35$ Cs$_{1-x}$Rb$_{x}$FeCl$_{3}$ undergoes a composition-driven transition from a gapped paramagnetic to a gapless and eventually magnetically ordered state. The transition itself, as well as the spin dynamics in its vicinity are strongly affected by disorder effects.

\section*{Acknowledgments}
This work was supported by Swiss National Science Foundation under Division 2.
We thank Dr. K. Yu. Povarov (ETH Z{\"u}rich) for assistance with the thermodynamics measurements and for fruitful discussion.
The neutron scattering experiment at the CNCS used resources at the Spallation Neutron Source, 
a DOE Office of Science User Facility operated by the Oak Ridge National Laboratory (IPTS-21713.1).


\end{document}